\documentclass{iopjournal}
\usepackage[T1]{fontenc}
\usepackage[utf8]{inputenc}
\usepackage{lmodern}
\usepackage{microtype}
\usepackage{amsmath,amssymb}
\usepackage{booktabs}
\usepackage{siunitx}
\usepackage{placeins}
\usepackage[round,authoryear]{natbib}
 \setcitestyle{aysep={},citesep={,}}
\graphicspath{{figures/}}
\makeatletter
\renewcommand\section{\@startsection{section}{1}{\z@}%
{-3.25ex\@plus -1ex \@minus -.2ex}%
{1.2ex \@plus .2ex}%
{\reset@font\normalsize\bfseries\raggedright}}
\renewcommand\subsection{\@startsection{subsection}{2}{\z@}%
{-3.25ex\@plus -1ex \@minus -.2ex}%
{0.8ex \@plus .2ex}%
{\reset@font\normalsize\itshape\raggedright}}
\renewcommand\subsubsection{\@startsection{subsubsection}{3}{\z@}%
{-3.25ex\@plus -1ex \@minus -.2ex}%
{0.8ex \@plus .2ex}%
{\reset@font\normalsize\itshape}}
\makeatother

\begin{document}

\title{Monte Carlo-based secondary dose check platform for the world's first ultra-compact upright
  proton therapy system}
  
\author{Toghrul Alakbarzade$^{1,2}$, Emir Musa Olgun$^{1,2}$, Yuan James Rao$^{3}$, Billy W Loo Jr$^{3}$,
  Adam Johansson$^{3}$, Yong Yang$^{3}$, Murat Surucu$^{3}$, Mehmet Burcin Unlu$^{2,4,5}$ and
  Serdar Charyyev$^{3,*}$}

\affil{$^1$Electrical and Electronics Engineering, Faculty of Engineering, Özyeğin University, İstanbul 34794, Türkiye}

\affil{$^2$ÖzU HealthTech Research Center, Özyeğin University, İstanbul 34794, Türkiye}

\affil{$^3$Department of Radiation Oncology, Stanford University School of Medicine, Palo Alto, CA 94305, United States of America}

\affil{$^4$Faculty of Engineering, Özyeğin University, İstanbul 34794, Türkiye}

\affil{$^5$Faculty of Aviation and Aeronautical Sciences, Özyeğin University, İstanbul 34794, Türkiye}

\affil{$^*$Author to whom any correspondence should be addressed.}

\email{charyyev@stanford.edu}


\begin{abstract}
\noindent\textbf{\textit{Objective.}} This work aimed to develop and validate an independent Monte Carlo (MC) secondary dose check of the RayStation treatment planning system (TPS) for the Mevion S250-FIT, the world's first ultra-compact upright proton therapy system.

\noindent\textbf{\textit{Approach.}} The treatment nozzle, including the energy modulation system of boron carbide and polycarbonate plates and the adaptive aperture, was modeled in the TOPAS (TOol for PArticle Simulation) MC code, and the beam model was commissioned against integrated depth dose
curves, in-air spot profiles, and dose per monitor unit measured at 12 energy settings.  Based on this model, an in-house converter, RayStation2TOPAS, recalculates the TPS plans by reconstructing each pencil beam spot, the aperture for each energy layer, and the seated patient's rotation. In total, 15 water phantom plans and 25 clinical plans were recalculated and compared with the TPS by the point dose, with a tolerance of \SI{5}{\percent}, and by three-dimensional gamma analysis at \SI{3}{\percent}/\SI{2}{mm}, with a pass rate of at least \SI{90}{\percent}.

\noindent\textbf{\textit{Main results.}} The simulated proton range agreed with the measurement within \SI{0.25}{mm}, the spot full width at half maximum within \SI{6.8}{\percent}, and the dose per monitor unit within \SI{1.3}{\percent}. The point dose differed from the TPS by \SI{-0.76}{\percent}
to \SI{+1.98}{\percent} in the water phantom plans and by \SI{-3.03}{\percent} to \SI{+3.99}{\percent} in the clinical plans, with gamma pass rates of \SI{96.39}{\percent} to \SI{99.93}{\percent} and \SI{90.08}{\percent} to \SI{99.98}{\percent}, so every plan passed both tests.

\noindent\textbf{\textit{Significance.}} To our knowledge, this is the first independent MC secondary dose check for the S250-FIT, and it complements measurement-based patient-specific quality assurance of its plans.
\end{abstract}

\section{Introduction}
The availability of proton therapy is limited by the size and cost of its facilities, much of which comes from the rotating gantry that delivers the beam to a supine patient from different angles \citep{Peeters2010, Yan2023}. Upright proton therapy replaces this rotating gantry with a fixed horizontal beamline and obtains the treatment angles by rotating the seated patient on a robotic positioner \citep{Mackie2021, Hegarty2022}. This design reduces both the footprint and the cost of the facility \citep{Volz2024}, and upright treatments on fixed beamlines are already in clinical use \citep{Gordon2021, Gordon2024, Lemaeva2024, Feldman2024, Blumenfeld2025, Blumenfeld2026}. The Mevion S250-FIT applies this approach in an ultra-compact system that combines a superconducting synchrocyclotron with an upright patient positioner and an upright CT scanner. The first unit was installed in a converted linear accelerator vault at Stanford and began treating patients in June 2026 \citep{Charyyev2026}.

As in any proton therapy system, before treatment, the dose calculated by the treatment planning system (TPS) for each plan is verified as part of patient-specific quality assurance, traditionally by measurements in a water phantom, which take beam time and test the calculated dose only in water \citep{Matter2018}. An independent secondary dose check,
which recalculates the plan in the patient geometry with a dose engine separate from that of the TPS, can replace these measurements \citep{Meijers2020}. Monte Carlo (MC) simulation is suited to this check because it is more accurate than analytical pencil beam algorithms in heterogeneous tissue \citep{Taylor2017}. For example, at the Maastro Proton Therapy Center, MC recalculation of more than 6000 plans over five years detected two planning errors that the measurements would have missed \citep{Rinaldi2026}.

However, an MC secondary dose check depends on the delivery system, because its beam model is
derived from the commissioning measurements of that system. Such checks have therefore been built for individual systems, including the Varian ProBeam \citep{Aitkenhead2020}, the IBA
Proteus ONE \citep{Liu2021}, and the Mevion S250i, for which MC models have been developed with a GPU-based \citep{Gajewski2020} and a CPU-based code \citep{Chiang2021}. The S250i is the gantry-mounted predecessor of the S250-FIT, but neither of its models describes the S250-FIT, whose energy modulation system replaces the 18 Lexan plates of the S250i with 10 boron carbide and 2 Lexan plates, which scatter the beam differently and give spots up to \SI{14}{\percent} smaller \citep{Charyyev2026}. At present, plans for the S250-FIT are calculated with the MC dose engine of RayStation, and to our knowledge, no independent MC dose check has been reported for this system.

To address this gap, this work presents an independent MC secondary dose check for the S250-FIT. First, the treatment nozzle was modeled in TOPAS, a CPU-based MC code \citep{Perl2012, Faddegon2020}, and the beam model was commissioned against the integrated depth dose curves, in-air spot profiles, and absolute dose per monitor unit measured at 12 energy settings during the clinical commissioning of the system \citep{Charyyev2026}. An in-house converter, RayStation2TOPAS, was then built on this commissioned model to recalculate RayStation plans. The converter reconstructs the delivery of each plan spot by spot, including the adaptive aperture, which can change shape between energy layers, and reproduces the upright treatment geometry, in which the beam is fixed and the seated patient is rotated and tilted. For each plan, the recalculated dose is compared with the RayStation dose at the point where RayStation reports the plan dose and by a three-dimensional gamma analysis. Finally, the check was validated on 15 water phantom plans and 25 clinical plans.

\section{Materials and methods}

\subsection{Mevion S250-FIT Upright Proton Therapy System}
\label{sec:system}

The system investigated in this work is a Mevion S250-FIT Upright Proton Therapy System (Mevion Medical Systems, Littleton, MA, USA), installed at the Stanford Cancer Center (Stanford FIT Proton Therapy). It is the world's first ultra-compact upright platform that delivers pencil beam scanning with a fixed horizontal beamline. Treatment angles are obtained by rotating the patient about the vertical axis on a six-degree-of-freedom upright patient positioner rather than rotating a massive gantry, and images are acquired by a diagnostic-quality upright fan-beam CT (Leo Cancer Care Inc., Middleton, WI, USA).

A compact superconducting synchrocyclotron, similar to that of its gantry-mounted predecessor, the S250i platform, accelerates protons to a fixed maximum energy of approximately \SI{227}{MeV}. Nested $x$--$y$ scanning coils steer the beam to each spot position at up to \SI{10}{mm/ms} before it enters the treatment nozzle. The treatment nozzle contains transmission ion chambers that monitor the beam, the energy modulation system (EMS), and the dynamic adaptive aperture (AA). The EMS has 12 plates, 2 polycarbonate (Lexan) and 10 boron carbide (B$_4$C), unlike the S250i, which uses 18 Lexan plates. The plates vary the beam energy over the range from about \SI{227}{MeV} to \SI{9}{MeV} with a specific plate configuration for each energy layer, and the switching time between energy layers is \SIrange{50}{500}{ms}. The AA sits downstream of the EMS and trims the beam laterally. It is a nickel mini-multileaf collimator with seven leaf pairs, each \SI{10}{cm} thick along the beam, carried on a translating stage that can position it at any point in the field. Its five central pairs are \SI{0.5}{cm} wide, and the two outer pairs, which act as jaws, are \SI{2}{cm} wide. The AA of the S250-FIT follows the same design as its predecessor. After the treatment nozzle, the beam passes through the exit window, which is a \SI{1.59}{mm} Lexan plate, before being delivered to a patient.

\subsection{Monte Carlo Model of the Treatment Nozzle}
\label{sec:model}

\begin{figure}[htbp]
\centering
\includegraphics[width=\textwidth]{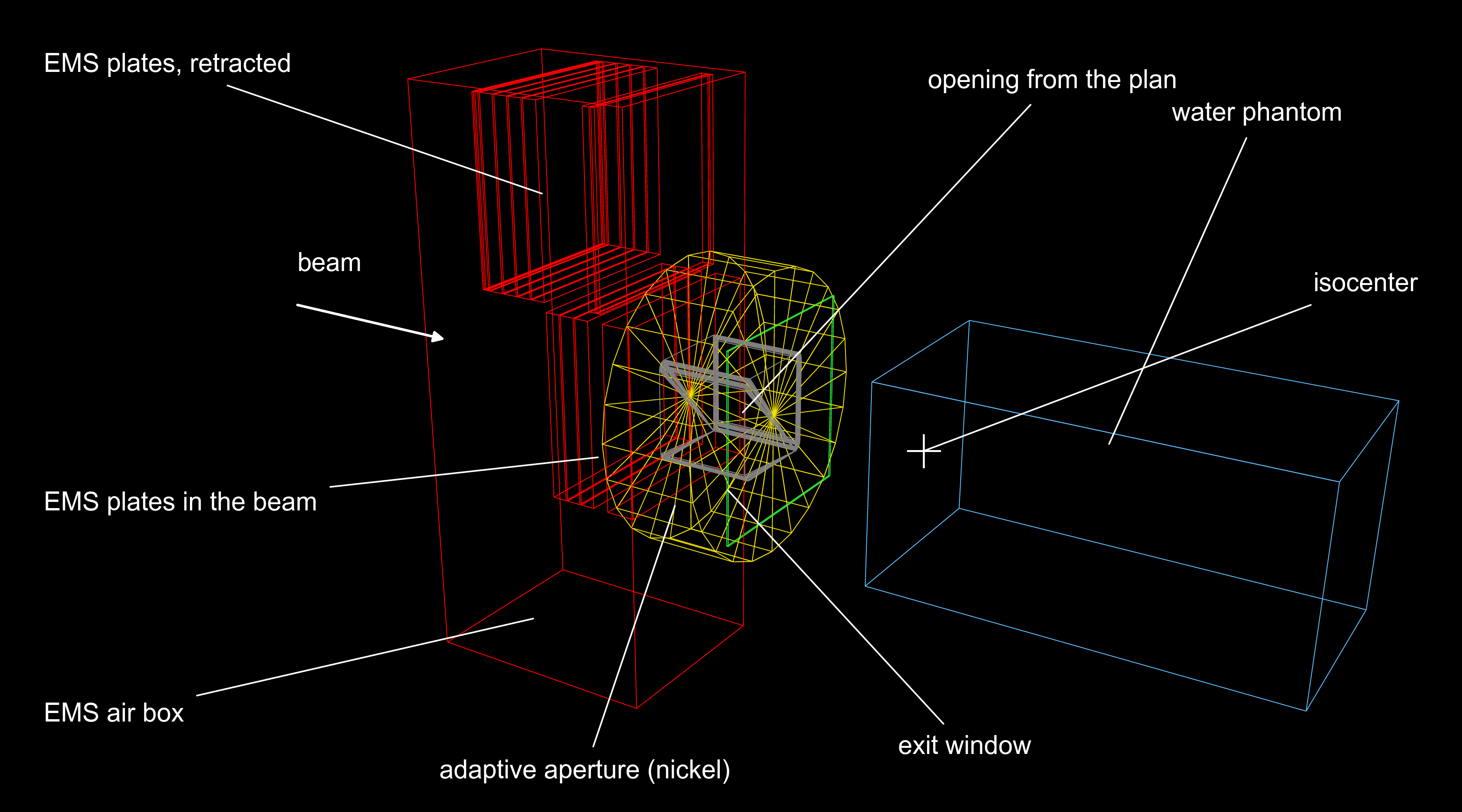}
\caption{The TOPAS model of the treatment nozzle, drawn by TOPAS from the input the converter wrote for one energy layer of the static plan of the \SI{1000}{\centi\metre\cubed} target at \SI{10}{cm} depth (section~\ref{sec:plans}), with the snout at \SI{200}{mm}. The arrow shows the direction of the beam, and the lines point from the labels to the components. The EMS plates (red) in this layer are in the beam, while the other plates are retracted, all inside the EMS air box. The AA (yellow) is a nickel cylinder with the opening of the plan (gray), followed by the exit window (green) and the water phantom (blue), whose surface is \SI{50}{mm} upstream of the isocenter (cross). The virtual source, \SI{1870}{mm} upstream of the isocenter, is outside the view.} 
\label{fig:nozzle}
\end{figure}

To develop a secondary dose check platform based on MC simulation, the treatment nozzle was modeled within an MC environment (figure~\ref{fig:nozzle}). All MC simulations were performed with TOPAS version 3.9 built on Geant4 10.07.p03 \citep{Agostinelli2003,Allison2006,Allison2016}. Electromagnetic interactions, including ionization energy loss and multiple Coulomb scattering, were
modeled with the \texttt{g4em-standard\_opt4} constructor. Inelastic nuclear interactions were modeled with \texttt{g4h-phy\_QGSP\_BIC\_HP}, and nuclear elastic scattering with
\texttt{g4h-elastic\_HP}. Production thresholds were set to \SI{0.05}{mm} for protons and \SI{0.5}{mm} for electrons, positrons, and photons.

The proton beam was modeled as a TOPAS emittance source with a bivariate Gaussian phase-space distribution on a plane \SI{1870}{mm} upstream of the isocenter. This distance is the virtual source-to-axis distance (SAD) of the scanning system, taken from the focal length of \SI{187}{cm} measured during the clinical commissioning of the system \citep{Charyyev2026}. The phase space can be specified with five parameters: the mean energy, relative energy spread, the spatial beam spread distribution (beam spot size), $\sigma$, the angular spread distribution (beam divergence), $\sigma'$, and the coefficient of correlation, $\rho$ \citep{Charyyev2020, Charyyev2025}. The same values were used for both transverse axes, consistent with the rotational symmetry of the measured in-air spots. Spot positioning was modeled as a rotation about the virtual source point as
\begin{equation}
    \theta_x = \arctan\!\left(\frac{x_s}{d_{\mathrm{SAD}}}\right), \qquad
    \theta_y = \arctan\!\left(\frac{y_s}{d_{\mathrm{SAD}}}\right),
\label{eq:steer}
\end{equation}
where $(x_s, y_s)$ is the planned spot position at the isocenter plane and $d_{\mathrm{SAD}} = \SI{1870}{mm}$. The EMS was modeled as 12 rectangular plates in air. Plate materials were assigned as the Geant4 reference materials for polycarbonate and boron carbide. The thickness of each boron carbide plate was set from its water-equivalent thickness in the system's range-shifter table. Each plate was placed either along the beam axis or laterally displaced from the field, depending on its state within the energy configuration.

For plans with open collimation, the AA leaves were retracted outside the scanned field. For static and dynamic collimation, the aperture was modeled as a nickel cylinder with a radius of \SI{150}{mm} (\texttt{TsAperture}) containing a straight-walled polygonal opening taken from the treatment plan. The representation for the collimated plans follows from what the treatment plan contains. The plan does not store the positions of the individual leaves; it stores the aperture opening as a polygon, a single one for static collimation and one per energy layer for dynamic collimation. Modeling the opening directly from this polygon reproduces the aperture as planned, without assuming how the machine converts the outline into leaf positions. The outer shape of the collimator does not affect the dose. Since \SI{100}{mm} of nickel stops protons of every deliverable energy, only the opening shapes the field, and the \SI{150}{mm} radius covers the entire scanned field at the aperture plane. Downstream of the AA, the exit window was modeled as a \SI{1.59}{mm} polycarbonate slab.

\subsection{Beam Model Commissioning and Benchmarking}
\label{sec:commissioning}

The TOPAS model was commissioned and tested against the clinical commissioning measurements of the S250-FIT. The same measurements were used to commission the beam model of the TPS, RayStation v2024A (RaySearch Laboratories, Stockholm, Sweden), and were exported from it. The exported file contains integrated depth dose (IDD) curves for pristine Bragg peaks, in-air spot profiles, and absolute dose per monitor unit (MU), together with the same quantities computed by the RayStation Monte Carlo (RS MC) dose engine. All three quantities were compared at the 12 energy settings of the clinical commissioning, from \SI{49.32}{MeV} to \SI{227.09}{MeV}. The setting at \SI{227.09}{MeV} is the maximum energy, delivered directly from the synchrocyclotron with no plate in the beam, and the other 11 settings were chosen so that each of the 12 EMS plates is in the beam in at least one of them. Each measurement was reproduced in TOPAS using the nozzle model from section~\ref{sec:model}. The mean energy and the energy spread were tuned against the IDD curves. The spot size, divergence, and correlation were obtained by a Bayesian fit to the RS MC in-air spot profiles at \SI{227.09}{MeV}.

\subsubsection{Integrated Depth Dose Curves}
\label{sec:idd}

A single spot was transported through the nozzle along the central axis, with the snout at \SI{200}{mm}, into a \SI{600}{mm} water cube, whose surface was at the isocenter. The
\texttt{DoseToWater} quantity was scored in a water-cylinder detector coaxial with the beam, \SI{600}{mm} long, divided into 1000 depth bins of \SI{0.6}{mm} each. Its diameter, \SI{81.6}{mm}, was set equal to that of the detector used in the measurement. The simulated IDDs were compared with the measured and RS MC curves after normalization to their maxima. For the comparison, the range $R_{80}$, the depth distal to the peak at which the dose falls to \SI{80}{\percent} of the maximum, the peak width $W_{80}$, the distance between the proximal and distal \SI{80}{\percent} points, and the distal fall-off $F_{80/20}$, the distance between the distal \SI{80}{\percent} and \SI{20}{\percent} points, were calculated.

\subsubsection{In-Air Spot Profiles}
\label{sec:profiles}
  
A single spot was transported through the nozzle along the central axis, with the snout at \SI{250}{mm}, into air. The \texttt{Fluence} quantity, restricted to protons, was scored in an air box centered on the isocenter, \SI{400}{mm} long and divided into \SI{1}{mm} bins along the beam and \SI{0.5}{mm} bins laterally, the same spacing as the measured profiles. Its lateral half-width, between 25 and \SI{85}{mm}, was set for each energy setting to ensure the box contained the entire spot. Profiles were taken in the five measurement planes, located at the isocenter and \SI{100}{mm} and \SI{200}{mm} upstream and downstream of it. In each plane, the profile along $x$ was the mean of the two \SI{0.5}{mm} rows of the \SI{1}{mm} slab on either side of the spot center, which form a line \SI{1}{mm} wide through the center, and the profile along $y$ was taken likewise. The full width at half maximum (FWHM) of each simulated profile was calculated from a two-Gaussian fit to the profile, and that of each measured and RS MC profile by linear interpolation at half of its maximum.

\subsubsection{Absolute Dose per MU}
\label{sec:dosemu}

The dose per MU was simulated under the conditions of the measurement. A uniform field of $100 \times \SI{100}{mm}$, made of $41 \times 41$ spots of equal weight with a spacing of \SI{2.5}{mm}, was delivered with the snout at \SI{200}{mm} into a water phantom of $201 \times 201 \times \SI{400}{mm}$, divided into \SI{3}{mm} bins laterally and \SI{1}{mm} bins in depth. The water surface was placed upstream of the isocenter by the reference depth, \SI{10}{mm} for energies up to \SI{114.52}{MeV} and \SI{50}{mm} for higher energies, so that the reference point was at the isocenter. The dose at the reference point was taken as the mean over the voxels within \SI{10}{mm} of the central axis, matching the \SI{20}{mm} diameter detector given in the commissioning data, and was interpolated linearly between the two depth bins around the reference depth.

TOPAS gives the dose per simulated proton, $d$, whereas the measurement gives the dose per MU, $D_{\mathrm{MU}}$. The two are related by the number of protons per MU, $N_{\mathrm{p/MU}}$:
\begin{equation}
    D_{\mathrm{MU}} = d \, N_{\mathrm{p/MU}}.
\label{eq:dosemu}
\end{equation}
The monitor chambers of the S250-FIT are upstream of the EMS \citep{Charyyev2026}, so they measure the protons before the plates slow them down, and $N_{\mathrm{p/MU}}$ is the same at every energy setting. A single value was therefore determined from the measurements, such that the simulated $D_{\mathrm{MU}}$ equals the measured $D_{\mathrm{MU}}$ on average over the 12 energy settings. With this value, the simulated $D_{\mathrm{MU}}$ at each energy setting was compared with the measured and RS MC values. The same $N_{\mathrm{p/MU}}$ converts the MU of the plans into protons in equation~(\ref{eq:dose}).

\subsection{RayStation2TOPAS}
\label{sec:rs2topas}

RayStation2TOPAS is an in-house converter that recalculates RayStation plans for the S250-FIT in TOPAS as an independent MC secondary dose check. The converter first parses the Digital Imaging and Communications in Medicine (DICOM) export of a plan into a list of spots, and then reconstructs the delivery and its absolute dose from that list in the nozzle model of section~\ref{sec:model}.

\subsubsection{Plan Parsing and Delivery Reconstruction}
\label{sec:delivery}

In the first stage, the converter parses the RT Ion Plan. The plan describes the delivery of each beam as a sequence of control points. The control points mark the progress of the delivery by the MU delivered so far, and each one gives the delivery parameters from that point on. Each energy layer is described by two control points, one at its start and one at its end. The start control point of each layer holds the EMS setting of the layer and the position and MU of every spot (``meterset weight'' in the plan). The first control point of a beam additionally holds the parameters of the whole beam: the isocenter, the snout position, and the rotation of the upright patient positioner (``patient support angle'' in the plan), together with the backrest tilt (``table-top pitch angle'' in the plan).

In the second stage, the converter turns the spot list into TOPAS runs. The converter groups the spots of each beam by EMS setting and simulates each group as a separate run. TOPAS steps through the spots in order, and at each step it aims the source at the spot and simulates a number of histories proportional to the spot's MU. The plan's total budget determines the number of histories per spot. The converter divides this budget among the runs in proportion to the MU of each run, the sum of the MU of its spots, so that run $c$ receives $N_c$ histories, at least one, and then divides $N_c$ among the spots of the run:
\begin{equation}
  N_s = \max\!\left(1,\ \mathrm{round}\!\left[N_c\,\frac{m_s}{\sum_{s'} m_{s'}}\right]\right),
\label{eq:hist}
\end{equation}
where $m_s$ is the MU of spot $s$ and the sum runs over the spots of run $c$. Rounding gives whole numbers of histories, and the lower limit of one history, for each run and for each spot, keeps every energy layer and every spot in the simulation.

Each run yields the dose for each simulated history. The converter obtains the absolute dose of the plan by scaling this dose to the number of protons the run represents and summing over the runs:
\begin{equation}
  D = n_{\mathrm{fx}} \sum_c \frac{D_c}{N_c}\, M_c\, N_{\mathrm{p/MU}},
\label{eq:dose}
\end{equation}
where $D_c$ is the dose of run $c$ and $N_c$ the number of histories simulated in it, so that $D_c/N_c$ is the dose per simulated proton. $M_c$ is the MU of run $c$ in one fraction, the sum of the MU of its spots, and $M_c N_{\mathrm{p/MU}}$ is the number of protons it delivers, with $N_{\mathrm{p/MU}}$ the number of protons per MU (section~\ref{sec:dosemu}). The sum over the runs therefore gives the dose of one fraction. It is multiplied by the number of planned fractions, $n_{\mathrm{fx}}$, because the plan gives the MU of a single fraction, whereas the exported TPS dose is for the entire treatment.
 
\subsubsection{Aperture Reconstruction}
\label{sec:aperture-recon}

How the aperture is modeled depends on the beam's collimation mode, which the converter determines from the plan. Open plans carry no aperture, and the AA is modeled with all leaves retracted (section~\ref{sec:model}). Static plans carry a single opening in the standard aperture block of the beam (\texttt{IonBlockSequence}), stored as a polygon, a list of $(x, y)$ vertices that the DICOM standard defines as
projected onto the isocenter plane \citep{DICOMPS33}, and the same opening is used for every energy layer. Dynamic plans reshape the AA between layers. RayStation stores the opening of each layer as a polygon of the same form, but in
private tags of the layer's start control point rather than in standard DICOM attributes. The converter reads these vertices and, for both static and dynamic plans, takes the material and thickness of the aperture from the aperture block.

The plan defines each opening at the isocenter plane, whereas the aperture sits upstream of it, so each polygon is scaled back along the diverging beam. Because the aperture walls are straight and the beam diverges, the downstream face of the aperture limits the field, and the polygon is scaled to that face by
\begin{equation}
  f = \frac{d_{\mathrm{SAD}} - |z_{\mathrm{d}}|}{d_{\mathrm{SAD}}},
\label{eq:scale}
\end{equation}
where $z_{\mathrm{d}}$ is the distance of the downstream face from the isocenter ($f = 0.887$ for a snout at \SI{200}{mm}).

\subsection{Validation Plans}
\label{sec:plans}

The converter was tested on 15 water phantom plans and 25 clinical plans. In the water phantom plans, five targets, a \SI{125}{\centi\metre\cubed} target at depths of 5, 10, and
\SI{20}{cm} and a \SI{1000}{\centi\metre\cubed} target at depths of 10 and \SI{20}{cm}, were each planned with open, static, and dynamic collimation. The plans used between 10 and 18 energy layers. Hence, the target depths and sizes tested the EMS model over different sets of plate configurations, and the three collimation modes tested the aperture reconstruction (section~\ref{sec:aperture-recon}). Each plan consisted of a single field with a patient support angle of \SI{270}{\degree}, the snout at \SI{200}{mm}, the isocenter at \SI{50}{mm} depth, and 30 fractions.

The clinical plans were the 25 plans of the first cohort of patients treated on the S250-FIT at Stanford. The plans covered 10 brain, 5 head and neck, 6 thoracic, 2 spine, and 2 bone treatment sites. Each patient was scanned seated in the treatment position on the upright fan-beam CT, with a slice thickness of \SI{2}{mm}, and was planned in RayStation v2024A with the MC proton dose engine, a statistical uncertainty of \SI{0.5}{\percent}, and a dose grid of \SI{2}{mm}. The plans had between 1 and 6 fields and backrest tilts between 0 and \SI{5}{\degree}. Static apertures were used in 14 plans and dynamic apertures in 11 plans, and the plans were delivered in 1 to 30 fractions. Unlike the water phantom plans, the clinical plans contained heterogeneous tissue, several fields at different patient support angles, and immobilization devices, so they also tested the CT conversion and the patient geometry (section~\ref{sec:dosecalc}).

\subsection{Dose Calculation}
\label{sec:dosecalc}

Water phantom plans were simulated in a water box of $201 \times 201 \times \SI{400}{mm}$, centered on the beam axis and divided into $67 \times 67 \times 400$ voxels of $3 \times 3 \times \SI{1}{mm}$, with the \SI{1}{mm} side along the beam, which is the voxel size of the RayStation dose grid. The box served as both the phantom and the scoring volume, so its front face was placed at the phantom surface given by the plan. The \texttt{DoseToWater} quantity was scored in each voxel.

Clinical plans were simulated in the planning CT, loaded in TOPAS as a DICOM patient (\texttt{TsDicomPatient}) and positioned so that the plan isocenter coincided with the isocenter of the nozzle model. On the S250-FIT, the beam is fixed, and the patient rotates, so in the simulation the patient was kept fixed and the whole nozzle, including the source, was rotated by the patient support angle of the beam about the vertical axis through the isocenter. When the plan included a backrest tilt, the patient was tilted by that angle about the left--right axis through the isocenter.
  
\begin{figure}[htbp]
\centering
\includegraphics[width=\textwidth]{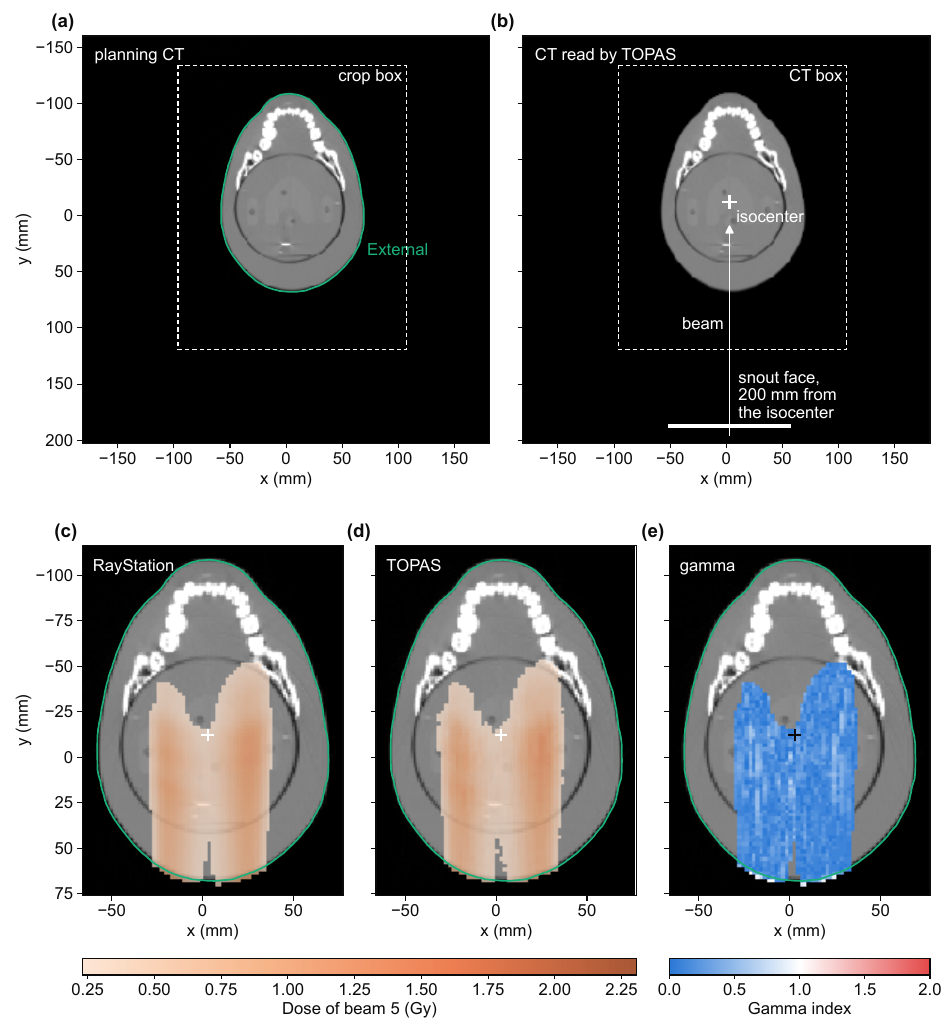}
\caption{The patient pipeline, shown on the CT of the IROC Houston anthropomorphic head-and-neck phantom, for one beam of its plan, on the axial slice through the isocenter. (a) Planning CT with the External outline and the box to which the CT was cropped. (b) The CT as TOPAS reads it, with the voxels outside the External set to air, the direction of the beam, and the snout face of the nozzle. (c) RayStation and (d) TOPAS dose of the beam, as physical dose, with TOPAS resampled onto the RayStation grid and set to zero outside the External. Dose below 10\,\% of the maximum is not shown. (e) Gamma index at 3\,\%/2\,mm. Because a single beam is shown, the dose-difference criterion and the low-dose threshold were percentages of the maximum of the RayStation dose of the beam, not of the dose at the isocenter as for a plan. The cross marks the isocenter.}
\label{fig:ctpipeline}
\end{figure}

In RayStation, the patient geometry used for the dose calculation consists of the volume inside
the External region of interest (ROI) and the ROIs of type Support and Fixation, and the rest of the volume is
treated as vacuum. The converter applied the same rule, and voxels outside these structures were set to air ($-1000$~HU). Where the plan overrides the material of a structure, TOPAS was given the RayStation material itself, with its elemental composition, mass density, and mean excitation energy. Where the plan gives only a density, the structure was given the CT number that yields that density in TOPAS. As in RayStation, the override of a Support or Fixation ROI applied to the whole ROI, whereas the override of any other ROI applied only to its part inside the External ROI. The CT was then cropped in-plane to a box around the patient and the support and fixation structures. Figure~\ref{fig:ctpipeline}(a) and (b) show these steps on the CT of the IROC Houston anthropomorphic head-and-neck phantom.

Each beam was simulated on its own copy of the CT. When the CT extended into the space of the nozzle, rows and columns containing only air were removed from the upstream side of the CT for that beam, so that no voxel values changed. The TOPAS overlap check was enabled for every simulation, and a simulation stopped if the patient still overlapped the nozzle.

CT numbers were converted to materials and mass densities with the stratified scheme of \citet{Schneider2000}, as implemented in TOPAS, with density correction factors derived from the clinical calibration curve of CT number to stopping-power ratio. The dose was scored in the voxels of the cropped CT itself, so the simulated dose had the same voxels as the CT. In each voxel, TOPAS converted the deposited energy to dose to water with the ratio of the stopping power of water to that of the voxel material (the
\texttt{DoseToWater} quantity), and the result was saved as a DICOM RT Dose file, the same format as the RayStation dose. The dose of each run was converted to absolute dose with equation~(\ref{eq:dose}), and the doses were summed on a common grid, since the upstream trim could give each beam a different grid.

\subsection{Dose Comparison}
\label{sec:compare}

The simulated dose was compared with the RayStation dose on the RayStation dose grid. It was transferred to that grid by volume-weighted resampling, in which each RayStation voxel received the mean of the simulated dose over its volume, weighted by its overlap with each simulation voxel. RayStation voxels covered by the simulation to less than \SI{99}{\percent} of their volume were excluded. For the clinical plans, the simulated dose was set to zero outside the volume in which RayStation computes dose, the External ROI together with the ROIs of type Support and Fixation. A voxel was kept when any part of it was within this volume.

Two tests were applied, following the structure of a calculation-based secondary dose check \citep{Zhu2021}, and both used the absolute dose. The first compared the dose at the point where RayStation reports the plan dose. This is the dose reference point of the plan, if it has one; otherwise, the beam dose specification point shared by its beams. In the water phantom plans, this point lies on the beam axis at the center of the spread-out Bragg peak (SOBP), and in the clinical plans at the isocenter. The mean over the $3 \times 3 \times 3$ voxels centered on it was compared, and a plan passed when the simulated dose agreed with the RayStation dose within \SI{5}{\percent}.

The second test compared the dose distributions with the three-dimensional gamma index \citep{Low1998}, computed with PyMedPhys \citep{Biggs2022} at \SI{3}{\percent}/\SI{2}{mm}, with a low-dose threshold of \SI{10}{\percent}, an interpolation fraction of 10, and a maximum gamma of 2. Neither distribution was rescaled, so the gamma test also detected an error in the overall dose level. Global normalization was used \citep{Miften2018}, so the dose-difference criterion and the low-dose threshold were expressed as percentages of a single reference dose across the whole plan. The reference dose was the RayStation dose, averaged over $3 \times 3 \times 3$ voxels, at the same point. Only a plan with neither point would use the maximum of the RayStation dose. A plan passed the gamma test when at least \SI{90}{\percent} of the evaluated voxels had a gamma index of 1 or less. Figure~\ref{fig:ctpipeline}(c)--(e) shows the two doses and the gamma index for one beam of the head-and-neck phantom plan.

\section{Results}
\label{sec:results}

\subsection{Beam Model}
\label{sec:res-beam}

\begin{figure}[htbp]
\centering
\includegraphics[width=\textwidth]{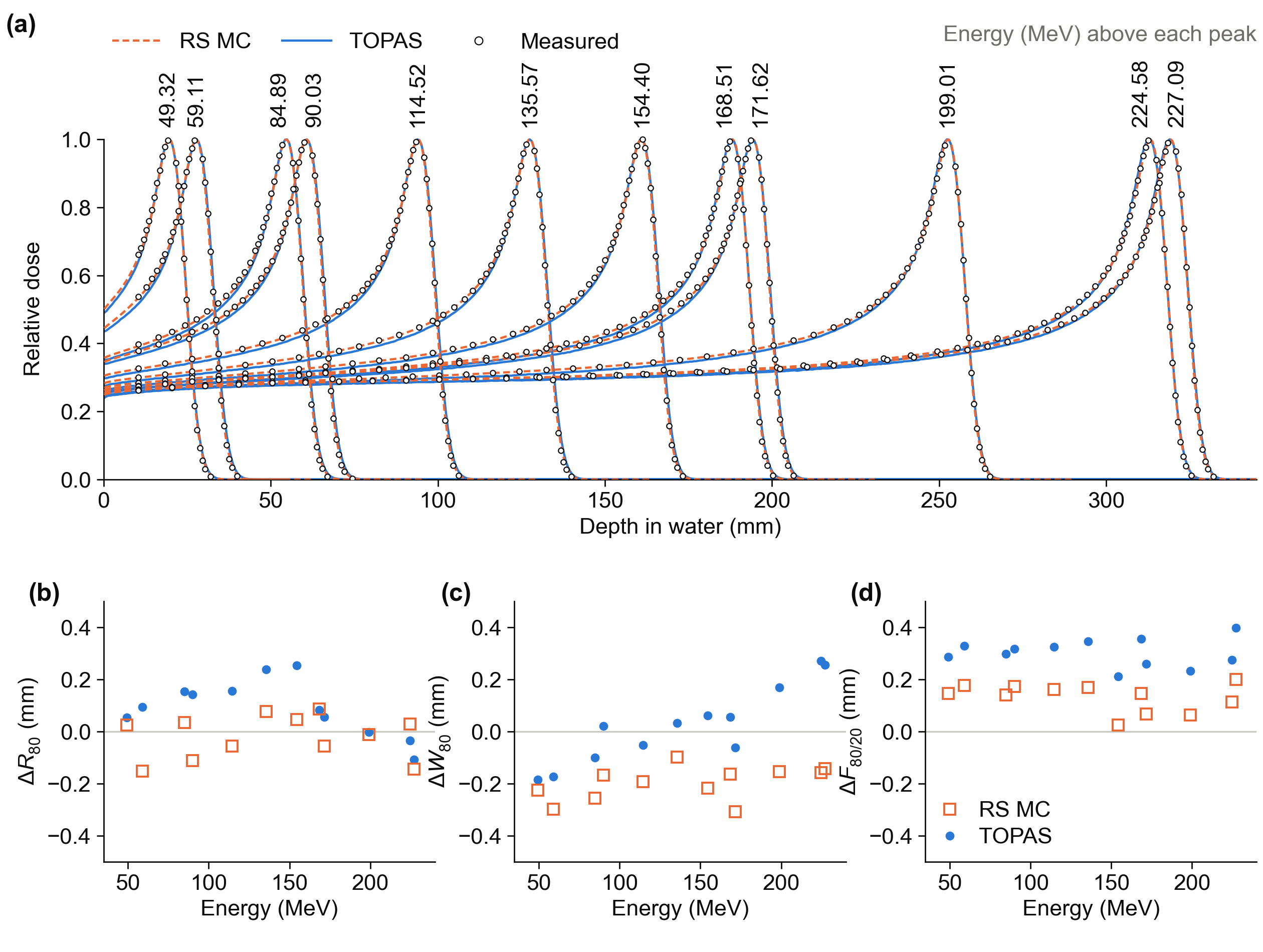}
\caption{Integrated depth dose curves at the 12 energy settings. (a) Simulated, measured, and RS MC curves, each normalized to its maximum. (b) to (d) Differences from the measurement in $R_{80}$, $W_{80}$, and $F_{80/20}$ for RS MC and TOPAS.}
\label{fig:idd}
\end{figure}

Table~\ref{tab:source} lists the source parameters obtained by tuning the model against the IDDs and the in-air spot profiles (section~\ref{sec:commissioning}). Table~\ref{tab:idd} and figure~\ref{fig:idd} compare the RS MC, simulated, and measured IDDs at 12 energy settings, from \SI{49.32}{MeV} to \SI{227.09}{MeV}. The simulated $R_{80}$ differed from the measurement by at most \SI{0.25}{mm}, at \SI{154.40}{MeV}, with a mean absolute difference of \SI{0.11}{mm}, while the RS MC $R_{80}$ differed by at most \SI{0.15}{mm}, at \SI{59.11}{MeV}, with a mean absolute difference of \SI{0.07}{mm}. The differences in the peak width $W_{80}$ ranged from \SI{-0.19}{mm} to \SI{+0.27}{mm} for TOPAS and from \SI{-0.31}{mm} to \SI{-0.10}{mm} for RS MC, and those in the distal fall-off $F_{80/20}$ from \SI{+0.21}{mm} to \SI{+0.40}{mm} and from \SI{+0.03}{mm} to \SI{+0.20}{mm}.

\begin{table}[htbp]
\centering
\caption{Source parameters of the beam model, identical for both transverse axes, defined at the virtual source plane \SI{1870}{mm} upstream of the isocenter.}
\label{tab:source}
\begin{tabular}{@{}llr@{}}
\toprule
Parameter & Symbol & Value \\
\midrule
Mean energy            & $E_0$        & \SI{228.12}{MeV} \\
Relative energy spread & $\Delta E/E$ & \SI{0.275}{\percent} \\
Spot size              & $\sigma$     & \SI{4.29}{mm} \\
Divergence             & $\sigma'$    & \SI{2.65}{mrad} \\
Correlation            & $\rho$       & $-0.816$ \\
\bottomrule
\end{tabular}
\end{table}

\begin{table}[htbp]
\centering
\caption{Measured range $R_{80}$ of the IDDs at the 12 energy settings, and the differences of RS MC and TOPAS from the measurement in the range $R_{80}$, the peak width $W_{80}$, and the distal fall-off $F_{80/20}$.}
\label{tab:idd}
\small
\begin{tabular}{@{}rrrrrrrr@{}}
\toprule
Energy & Measured $R_{80}$ & \multicolumn{2}{c}{$\Delta R_{80}$ (mm)} & \multicolumn{2}{c}{$\Delta W_{80}$ (mm)}
& \multicolumn{2}{c}{$\Delta F_{80/20}$ (mm)} \\
\cmidrule(lr){3-4}\cmidrule(lr){5-6}\cmidrule(l){7-8}
(MeV) & (mm) & RS MC & TOPAS & RS MC & TOPAS & RS MC & TOPAS \\
\midrule
49.32 &  22.88 & $+0.02$ & $+0.05$ & $-0.23$ & $-0.19$ & $+0.15$ & $+0.29$ \\
59.11 &  31.04 & $-0.15$ & $+0.09$ & $-0.30$ & $-0.17$ & $+0.18$ & $+0.33$ \\
84.89 &  57.87 & $+0.03$ & $+0.15$ & $-0.26$ & $-0.10$ & $+0.14$ & $+0.30$ \\
90.03 &  64.08 & $-0.11$ & $+0.14$ & $-0.17$ & $+0.02$ & $+0.17$ & $+0.32$ \\
114.52 &  97.39 & $-0.06$ & $+0.16$ & $-0.19$ & $-0.05$ & $+0.16$ & $+0.32$ \\
135.57 & 130.72 & $+0.08$ & $+0.24$ & $-0.10$ & $+0.03$ & $+0.17$ & $+0.35$ \\
154.40 & 164.10 & $+0.05$ & $+0.25$ & $-0.22$ & $+0.06$ & $+0.03$ & $+0.21$ \\
168.51 & 191.34 & $+0.09$ & $+0.08$ & $-0.16$ & $+0.06$ & $+0.15$ & $+0.36$ \\
171.62 & 197.56 & $-0.06$ & $+0.06$ & $-0.31$ & $-0.06$ & $+0.07$ & $+0.26$ \\
199.01 & 255.99 & $-0.01$ & $0.00$ & $-0.15$ & $+0.17$ & $+0.06$ & $+0.23$ \\
224.58 & 316.45 & $+0.03$ & $-0.04$ & $-0.16$ & $+0.27$ & $+0.11$ & $+0.28$ \\
227.09 & 322.71 & $-0.14$ & $-0.11$ & $-0.14$ & $+0.26$ & $+0.20$ & $+0.40$ \\
\bottomrule
\end{tabular}
\end{table}

\begin{figure}[htbp]
\centering
\includegraphics[width=\textwidth]{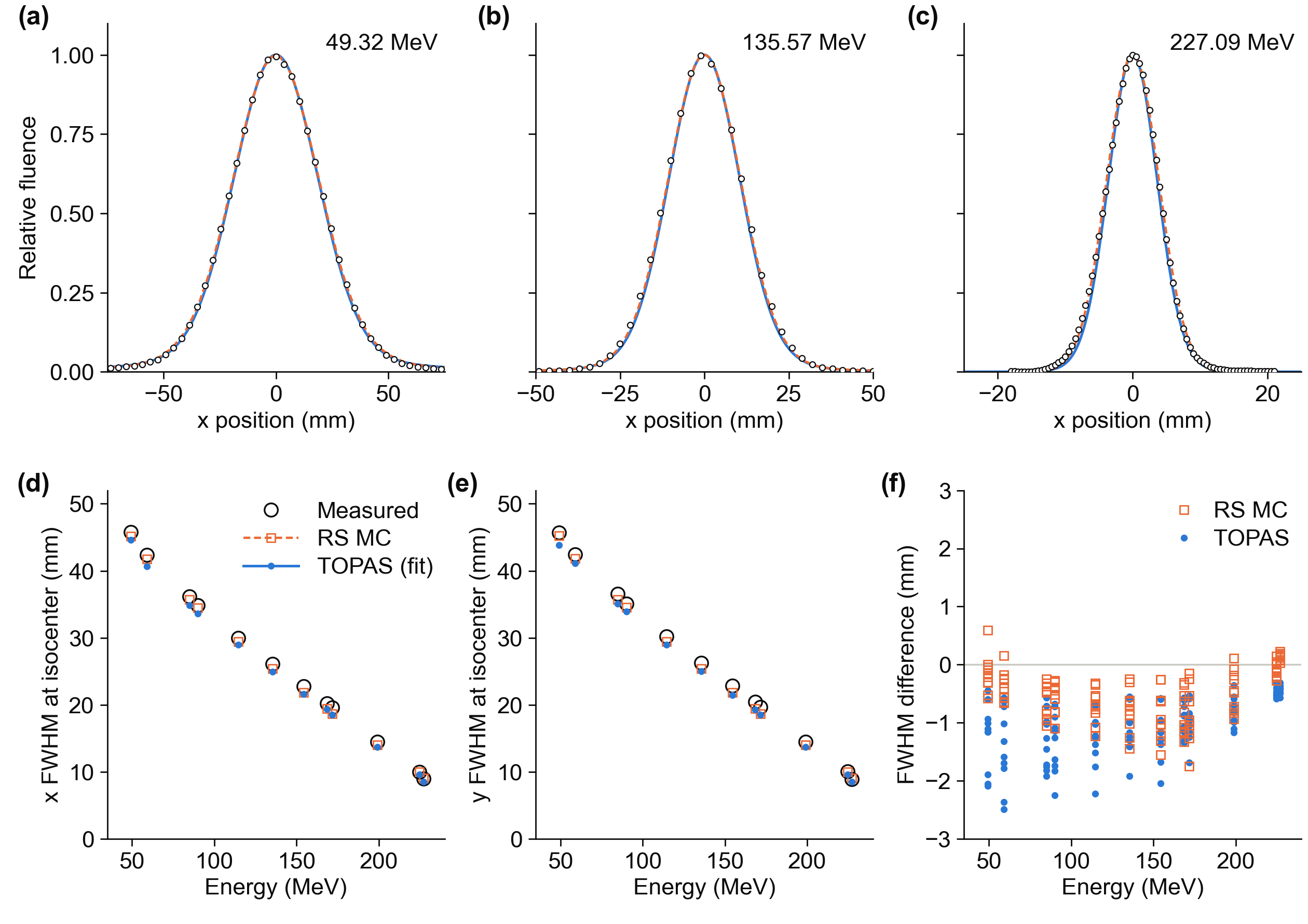}
\caption{In-air spot profiles and widths. (a) to (c) Simulated, measured, and RS MC profiles along $x$ at the isocenter plane at \SI{49.32}{MeV}, \SI{135.57}{MeV}, and \SI{227.09}{MeV}. The simulated profile is the two-Gaussian fit to the proton fluence along a line through the spot center, normalized to its maximum; the measured and RS MC profiles are normalized to their maxima. Each profile is centered on the midpoint of its FWHM. (d, e) FWHM at the isocenter plane along $x$ and $y$ at the 12 energy settings. (f) FWHM difference from the measurement, RS MC or TOPAS minus measured, over all five planes and both axes, 120 comparisons each.}
\label{fig:spot}
\end{figure}

Table~\ref{tab:fwhm} and figure~\ref{fig:spot} compare the RS MC, simulated, and measured spot FWHM at the same 12 settings. Over all five planes and both axes, 120 comparisons in total, the TOPAS FWHM differed from the measured FWHM by at most \SI{2.49}{mm}, at \SI{59.11}{MeV}, with a mean absolute difference of \SI{1.05}{mm}, while the RS MC FWHM differed by at most \SI{1.75}{mm}, at \SI{171.62}{MeV}, with a mean absolute difference of \SI{0.56}{mm}. The largest relative differences were \SI{6.8}{\percent} for TOPAS and \SI{6.2}{\percent} for RS MC.

\begin{table}[htbp]
\centering
\caption{Measured FWHM of the in-air spots at the isocenter plane on both axes at the 12 energy settings, and the differences of RS MC and TOPAS from the measurement.}
\label{tab:fwhm}
\small
\begin{tabular}{@{}rrrrrrr@{}}
\toprule
Energy & \multicolumn{3}{c}{$x$ FWHM (mm)} & \multicolumn{3}{c}{$y$ FWHM (mm)} \\
\cmidrule(lr){2-4}\cmidrule(l){5-7}
(MeV) & Measured & $\Delta$ RS MC & $\Delta$ TOPAS & Measured & $\Delta$ RS MC & $\Delta$ TOPAS \\
\midrule
49.32 & 45.75 & $-0.57$ & $-1.15$ & 45.73 & $-0.52$ & $-1.89$ \\
59.11 & 42.35 & $-0.57$ & $-1.70$ & 42.46 & $-0.64$ & $-1.32$ \\
84.89 & 36.16 & $-0.48$ & $-1.26$ & 36.56 & $-0.85$ & $-1.45$ \\
90.03 & 34.86 & $-0.39$ & $-1.26$ & 35.10 & $-0.58$ & $-1.14$ \\
114.52 & 29.99 & $-0.56$ & $-1.02$ & 30.21 & $-0.76$ & $-1.25$ \\
135.57 & 26.10 & $-0.69$ & $-1.16$ & 26.25 & $-0.86$ & $-1.24$ \\
154.40 & 22.78 & $-0.94$ & $-1.16$ & 22.85 & $-1.03$ & $-1.37$ \\
168.51 & 20.25 & $-0.82$ & $-0.88$ & 20.47 & $-1.03$ & $-1.15$ \\
171.62 & 19.61 & $-0.95$ & $-1.05$ & 19.69 & $-1.04$ & $-1.16$ \\
199.01 & 14.51 & $-0.45$ & $-0.83$ & 14.53 & $-0.48$ & $-0.79$ \\
224.58 & 10.00 & $+0.02$ & $-0.40$ & 10.11 & $-0.10$ & $-0.50$ \\
227.09 &  8.99 & $+0.12$ & $-0.47$ &  8.93 & $+0.16$ & $-0.43$ \\
\bottomrule
\end{tabular}
\end{table}

\begin{figure}[htbp]
\centering
\includegraphics[width=\textwidth]{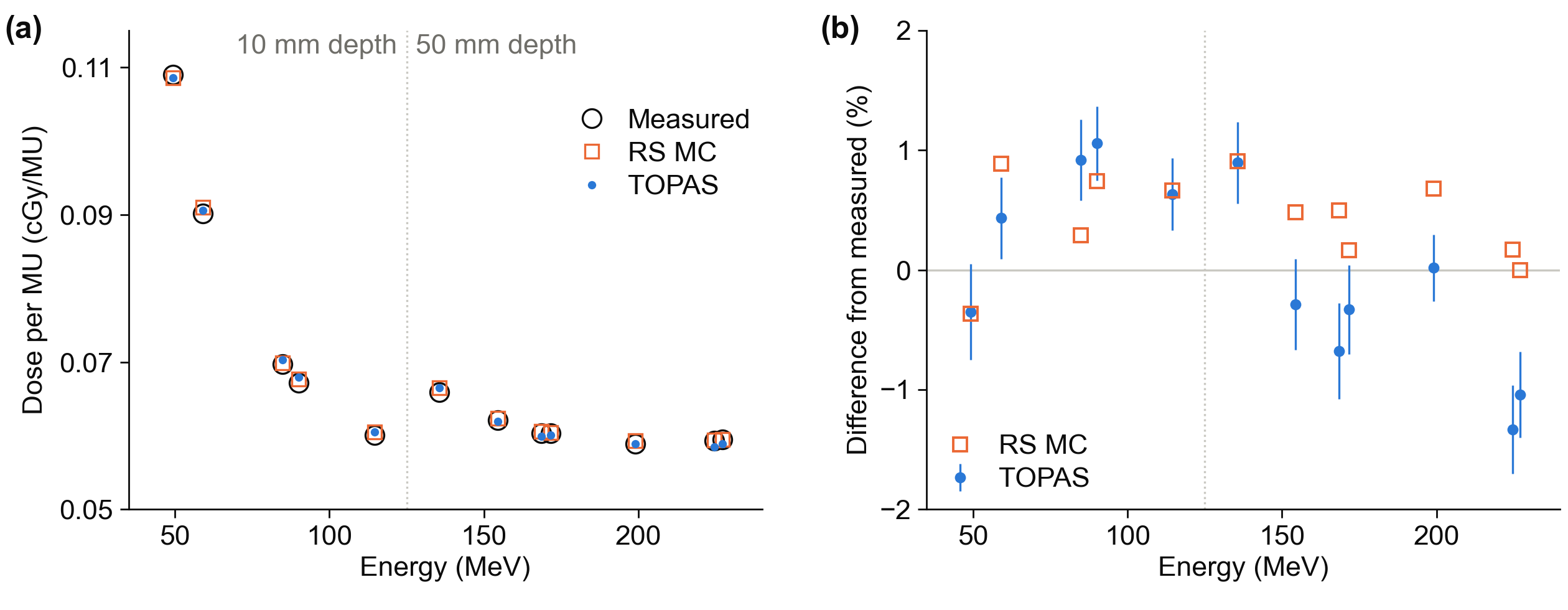}
\caption{Dose per MU at the calibration condition. (a) Simulated, measured, and RS MC dose per MU at the 12 energy settings. The dotted line separates the settings with a reference depth of \SI{10}{mm} from those with \SI{50}{mm}. (b) Difference from the measurement for RS MC and TOPAS. The mean TOPAS difference is zero by construction, because the number of protons per MU was set from these 12 settings (section~\ref{sec:dosemu}). Error bars are the statistical uncertainty of the simulated dose.}
\label{fig:dosemu}
\end{figure}

Table~\ref{tab:dosemu} and figure~\ref{fig:dosemu} compare the RS MC, simulated, and measured dose per MU at the same 12 settings. The number of protons per MU was $7.896 \times 10^{7}$, and the numbers that reproduced the measurement at each energy setting ranged from $7.81 \times 10^{7}$ to $8.00 \times 10^{7}$. The simulated dose per MU differed from the measurement by \SI{-1.3}{\percent} to \SI{+1.1}{\percent}, with a mean absolute difference of \SI{0.7}{\percent}, while the RS MC dose per MU differed by \SI{-0.4}{\percent} to \SI{+0.9}{\percent}, with a mean absolute difference of \SI{0.5}{\percent}.

\begin{table}[htbp]
\centering
\caption{Dose per MU at the calibration condition: $100 \times \SI{100}{mm}$ field, \SI{2.5}{mm} spot spacing, mean over a \SI{20}{mm} diameter region at the reference depth, with $7.896 \times 10^{7}$ protons per MU. Measured and RS MC values are from the clinical commissioning data.}
\label{tab:dosemu}
\small
\begin{tabular}{@{}rrrrrrr@{}}
\toprule
Energy & Depth & Measured & RS MC & TOPAS & \multicolumn{2}{c}{Difference from measured (\%)} \\
\cmidrule(l){6-7}
(MeV) & (mm) & \multicolumn{3}{c}{(cGy/MU)} & RS MC & TOPAS \\
\midrule
49.32 & 10 & 0.1090 & 0.1086 & 0.10862 & $-0.4$ & $-0.3$ \\
59.11 & 10 & 0.0902 & 0.0910 & 0.09059 & $+0.9$ & $+0.4$ \\
84.89 & 10 & 0.0697 & 0.0699 & 0.07034 & $+0.3$ & $+0.9$ \\
90.03 & 10 & 0.0672 & 0.0677 & 0.06791 & $+0.7$ & $+1.1$ \\
114.52 & 10 & 0.0601 & 0.0605 & 0.06048 & $+0.7$ & $+0.6$ \\
135.57 & 50 & 0.0659 & 0.0665 & 0.06649 & $+0.9$ & $+0.9$ \\
154.40 & 50 & 0.0621 & 0.0624 & 0.06192 & $+0.5$ & $-0.3$ \\
168.51 & 50 & 0.0603 & 0.0606 & 0.05989 & $+0.5$ & $-0.7$ \\
171.62 & 50 & 0.0603 & 0.0604 & 0.06010 & $+0.2$ & $-0.3$ \\
199.01 & 50 & 0.0589 & 0.0593 & 0.05891 & $+0.7$ & $0.0$ \\
224.58 & 50 & 0.0593 & 0.0594 & 0.05851 & $+0.2$ & $-1.3$ \\
227.09 & 50 & 0.0595 & 0.0595 & 0.05888 & $0.0$ & $-1.0$ \\
\bottomrule
\end{tabular}
\end{table}

\subsection{Water Phantom Plans}
\label{sec:res-water}

\begin{figure}[htbp]
\centering
\includegraphics[width=\textwidth]{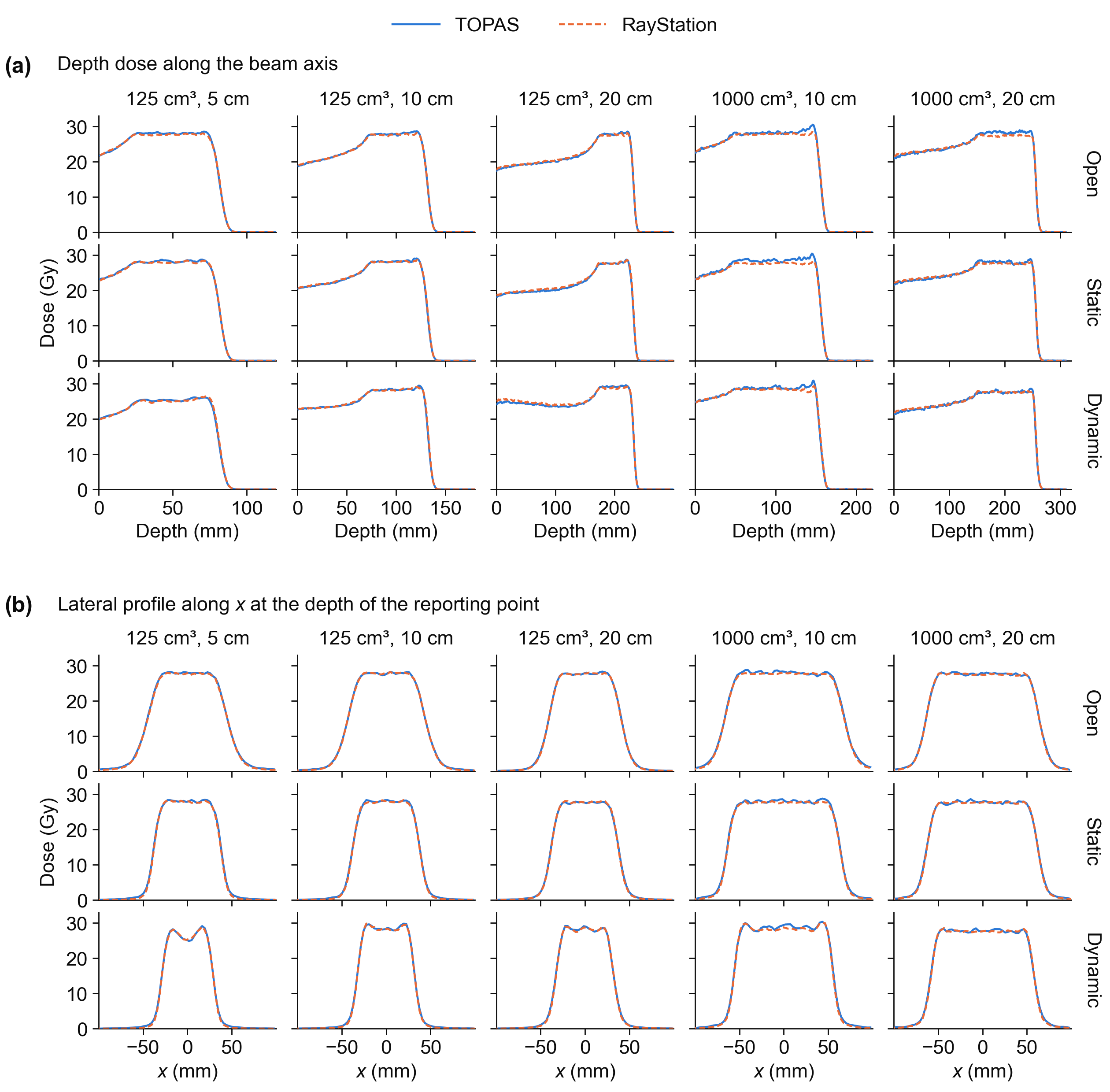}
\caption{RayStation and TOPAS dose of the 15 water phantom plans, as physical dose summed over the 30 fractions on the RayStation dose grid, one panel per plan. Columns are the targets and rows the collimation modes. (a) Depth dose along the beam axis. (b) Lateral profile along $x$ at the depth of the point where RayStation reports the plan dose, through the voxel nearest to that point. Solid lines are TOPAS and dashed lines RayStation.}
\label{fig:water}
\end{figure}

Table~\ref{tab:waterres} lists the point dose and gamma results of the 15 water phantom plans, and figure~\ref{fig:water} shows their depth doses and lateral profiles. The simulated point dose differed from the RayStation dose by $-0.76$ to $+1.98$\,\%, with a mean of $+0.64$\,\%, and all 15 plans passed the point dose test. The largest difference, $+1.98$\,\%, was found for the static plan of the \SI{1000}{\centi\metre\cubed} target at \SI{10}{cm}. The gamma pass rate ranged from 96.39 to 99.93\,\%, with a mean of 98.94\,\%, and all 15 plans passed the gamma test.

For the \SI{125}{\centi\metre\cubed} target, the point dose difference ranged from $-0.76$ to $+1.05$\,\% (mean $+0.27$\,\%) and the gamma pass rate from 99.39 to 99.93\,\% (mean 99.67\,\%). For the \SI{1000}{\centi\metre\cubed} target, the point dose difference ranged from $+0.66$ to $+1.98$\,\% (mean $+1.18$\,\%) and the gamma pass rate from 96.39 to 99.04\,\% (mean 97.85\,\%). The mean point dose difference was $+0.77$\,\% for the open plans, $+0.77$\,\% for the static plans, and $+0.36$\,\% for the dynamic plans, and the mean gamma pass rate was 99.03, 98.81, and 98.99\,\%, respectively.

\begin{table}[htbp]
\centering
\caption{Point dose and gamma results of the water phantom plans. Doses are physical doses summed over the 30 fractions, at the point where RayStation reports the plan dose. The gamma pass rate is for 3\,\%/2\,mm on absolute dose with global normalization.}
\label{tab:waterres}
\small
\setlength{\tabcolsep}{4pt}
\begin{tabular}{rrlrrrrrrr}
\toprule
Target & Depth & Collimation & Layers & Spots & RayStation & TOPAS & Difference & Gamma \\
(\si{\centi\metre\cubed}) & (cm) & & & & (cGy) & (cGy) & (\%) & (\%) \\
\midrule
125  & 5  & Open    & 10 & 900  & 2775.84 & 2805.05 & $+1.05$ & 99.47 \\
125  & 5  & Static  & 10 & 686  & 2786.93 & 2809.88 & $+0.82$ & 99.40 \\
125  & 5  & Dynamic & 14 & 503  & 2553.02 & 2533.69 & $-0.76$ & 99.93 \\
125  & 10 & Open    & 10 & 937  & 2774.11 & 2777.39 & $+0.12$ & 99.51 \\
125  & 10 & Static  & 10 & 789  & 2802.71 & 2817.93 & $+0.54$ & 99.39 \\
125  & 10 & Dynamic & 10 & 304  & 2825.09 & 2834.39 & $+0.33$ & 99.71 \\
125  & 20 & Open    & 10 & 951  & 2763.53 & 2773.12 & $+0.35$ & 99.91 \\
125  & 20 & Static  & 10 & 935  & 2772.48 & 2768.98 & $-0.13$ & 99.89 \\
125  & 20 & Dynamic & 10 & 454  & 2844.37 & 2847.93 & $+0.12$ & 99.82 \\
1000 & 10 & Open    & 18 & 3729 & 2773.86 & 2809.89 & $+1.30$ & 97.23 \\
1000 & 10 & Static  & 18 & 3475 & 2759.83 & 2814.53 & $+1.98$ & 96.39 \\
1000 & 10 & Dynamic & 18 & 1974 & 2840.60 & 2876.37 & $+1.26$ & 96.79 \\
1000 & 20 & Open    & 18 & 3658 & 2745.78 & 2774.40 & $+1.04$ & 99.04 \\
1000 & 20 & Static  & 18 & 3643 & 2761.01 & 2779.33 & $+0.66$ & 98.98 \\
1000 & 20 & Dynamic & 18 & 2951 & 2755.56 & 2779.28 & $+0.86$ & 98.68 \\
\bottomrule
\end{tabular}
\end{table}

\subsection{Clinical Plans}
\label{sec:res-patients}

Table~\ref{tab:patres} compares the simulated and RayStation doses of the 25 clinical plans. The point dose differed from the RayStation dose by \SI{-3.03}{\percent} to \SI{+3.99}{\percent}, with a mean of \SI{+0.63}{\percent}, and all 25 plans passed the point dose test. The largest difference, \SI{+3.99}{\percent}, was for plan 10. The gamma pass rate ranged from \SI{90.08}{\percent} to \SI{99.98}{\percent}, and all 25 plans passed the gamma test.

For the 14 plans with static apertures, the point dose difference ranged from \SI{-2.09}{\percent} to \SI{+2.12}{\percent} and the gamma pass rate from \SI{90.08}{\percent} to \SI{99.98}{\percent}. For the 11 plans with dynamic apertures, the point dose difference ranged from \SI{-3.03}{\percent} to \SI{+3.99}{\percent} and the gamma pass rate from \SI{90.99}{\percent} to \SI{99.97}{\percent}.

\begin{table}[htbp]
\centering
\caption{Point dose and gamma results of the clinical plans. Doses are physical doses for the whole plan. Differences are simulation minus RayStation, relative to RayStation, computed before rounding. Time is the time of the whole check of the plan, including the preparation of the simulation, on a workstation with two Intel Xeon Gold 6248R processors (48 cores, \SI{3.0}{GHz}) and \SI{376}{GB} of memory.} 
\label{tab:patres}
\small
\setlength{\tabcolsep}{4pt}
\begin{tabular}{@{}lllrrrrrrr@{}}
\toprule
& & & & & \multicolumn{2}{c}{Point dose (cGy)} & Difference & Gamma & Time \\
\cmidrule(lr){6-7}
Plan & Site & Aperture & Fields & Fractions & RayStation & TOPAS & (\%) & (\%) & (min) \\
\midrule
1  & Neck          & Static  & 2 & 15 & 2079.9 & 2067.5 & $-0.60$ & 99.20 & 115 \\
2  & Mandible      & Static  & 4 & 30 & 5555.3 & 5535.0 & $-0.37$ & 99.64 & 118 \\
3  & Thorax        & Static  & 5 &  9 & 2604.2 & 2648.2 & $+1.69$ & 99.30 & 169 \\
4  & Skull base    & Static  & 4 &  5 & 1906.4 & 1890.8 & $-0.82$ & 99.50 & 202 \\
5  & Brain         & Static  & 4 & 30 & 6136.7 & 6208.2 & $+1.17$ & 99.63 & 166 \\
6  & Brain         & Static  & 3 &  6 & 1015.0 &  993.7 & $-2.09$ & 98.70 & 150 \\
7  & Bone          & Static  & 2 &  1 &  744.6 &  756.6 & $+1.60$ & 97.81 & 101 \\
8  & Brain         & Static  & 4 & 30 & 4977.8 & 5030.5 & $+1.06$ & 99.98 & 249 \\
9  & Brain         & Dynamic & 3 & 20 & 3345.6 & 3414.2 & $+2.05$ & 99.97 & 123 \\
10 & Thorax        & Dynamic & 3 & 15 & 5617.4 & 5841.3 & $+3.99$ & 94.08 & 120 \\
11 & Chest wall    & Dynamic & 3 &  5 & 2843.8 & 2866.5 & $+0.80$ & 97.41 & 136 \\
12 & Brain         & Dynamic & 4 & 17 & 3558.8 & 3666.6 & $+3.03$ & 90.99 & 173 \\
13 & Spine         & Static  & 2 & 27 & 4502.5 & 4516.0 & $+0.30$ & 95.75 & 87 \\
14 & Brain         & Dynamic & 4 & 17 & 2909.1 & 2863.1 & $-1.58$ & 99.75 & 157 \\
15 & Thorax        & Static  & 3 &  8 & 1584.4 & 1602.5 & $+1.14$ & 99.60 & 111 \\
16 & Head and neck & Dynamic & 4 & 25 & 4680.0 & 4789.5 & $+2.34$ & 92.08 & 179 \\
17 & Head and neck & Dynamic & 4 & 19 & 3551.3 & 3443.7 & $-3.03$ & 93.01 & 200 \\
18 & Brain         & Dynamic & 4 & 30 & 6443.0 & 6349.9 & $-1.45$ & 99.53 & 129 \\
19 & Lung          & Static  & 1 &  5 & 1718.4 & 1754.8 & $+2.12$ & 99.36 & 55 \\
20 & Lung          & Dynamic & 3 &  4 & 4420.7 & 4321.5 & $-2.24$ & 99.42 & 87 \\
21 & Rib           & Static  & 2 &  5 & 3374.2 & 3428.8 & $+1.62$ & 99.90 & 68 \\
22 & Brain         & Dynamic & 6 & 30 & 6187.6 & 6383.3 & $+3.16$ & 97.98 & 187 \\
23 & Spine         & Static  & 2 &  1 &  178.0 &  180.6 & $+1.46$ & 90.08 & 92 \\
24 & Brain         & Dynamic & 5 & 10 & 1662.5 & 1661.9 & $-0.03$ & 99.71 & 190 \\
25 & Head and neck & Static  & 3 & 20 & 3809.8 & 3828.8 & $+0.50$ & 99.63 & 116 \\

\bottomrule
\end{tabular}
\end{table}

\section{Discussion}

This work developed an independent MC secondary dose check for the Mevion S250-FIT. The TOPAS model of the nozzle agreed with the IDDs, in-air spot profiles, and dose per MU measured at all 12 energy settings during the clinical commissioning. Built on this model, RayStation2TOPAS recalculated 15 water phantom plans and 25 clinical plans. The water phantom plans covered open, static, and dynamic collimation, and the clinical plans added heterogeneous patient anatomy and the upright treatment geometry. Every plan passed both the point dose and the gamma test, so the check can be used for patient-specific quality assurance of the S250-FIT, as an independent MC check has been used for five years for the S250i at the Maastro Proton Therapy Center \citep{Rinaldi2026}. The remaining differences between TOPAS, the measurements, and RayStation, and the limitations of the check, are discussed below.

The range $R_{80}$ of the simulated IDDs agreed with the measurement within \SI{0.25}{mm} at all 12 energy settings. The spot FWHM agreed with the measurement within \SI{6.8}{\percent} in all 120 comparisons, and that of the RS MC model of the clinical TPS within \SI{6.2}{\percent}. This agreement is similar to that reported for the TOPAS model of the S250i, whose IDDs agreed with the measurements within \SI{1}{mm} and whose spot sizes agreed within \SI{10}{\percent} \citep{Chiang2021}. The simulated dose per MU agreed with the measurement within \SI{1.3}{\percent} at every energy setting, and that of RS MC within \SI{0.9}{\percent}.
      
The water phantom plans tested the beam model and the reconstruction of the delivery, including the adaptive aperture, in a homogeneous medium. In these plans, open, static, and dynamic collimation agreed with RayStation with nearly the same mean gamma pass rate, \SI{99.03}{\percent}, \SI{98.81}{\percent}, and \SI{98.99}{\percent}, respectively. The clinical plans added the patient anatomy and the upright geometry to the check. The planning CT, with the material overrides of RayStation, was converted to materials and mass densities, and each field was simulated with the rotation of the seated patient. With these additional steps, the point dose differences spread more widely than in the water phantom plans, from $-3.03$ to $+3.99$\,\%. The spread may also come
from the reporting point. In the water phantom plans, this point was the center of the SOBP, where the dose is uniform, whereas in the clinical plans it was the isocenter, where the dose need not be uniform. Even the largest difference was within the \SI{5}{\percent} tolerance, and every plan passed the gamma test.

The check has a few limitations. First, it verifies the dose calculation of RayStation, but not the transfer of the plan to the treatment machine or its delivery. Reconstructing the delivered dose from the machine log files, as done at other centers \citep{Matter2018, Rinaldi2026}, would extend the check to these steps. Second, the shape of the adaptive aperture is taken from the outline planned in RayStation, without modeling how the machine converts this outline into leaf positions (section~\ref{sec:aperture-recon}). All 25 clinical plans passed the check with this approach, but its long-term effect is not yet known and should be monitored as more plans are checked. Finally, checking a clinical plan took 55 to 249 min on a CPU workstation. This time would be long for a center that starts several new patients per day, but one or two new patients per week
start treatment on the S250-FIT at Stanford, so each plan can be checked before its first treatment fraction. Most of this time was spent preparing the plan for the simulation rather than in TOPAS itself, so optimizing this preparation would shorten the check in future applications.

\section{Conclusion}

An independent MC secondary dose check based on TOPAS was developed and validated for the Mevion S250-FIT. Its model of the nozzle agreed with the IDDs, in-air spot profiles, and dose per MU measured during the clinical commissioning at all 12 energy settings. Built on this model, RayStation2TOPAS recalculates RayStation plans, including the adaptive aperture of each energy layer and the upright treatment geometry. In the 15 water phantom plans and 25 clinical plans, the recalculated point dose differed from the RayStation dose by at most \SI{2.0}{\percent} and \SI{4.0}{\percent}, the gamma pass rate at \SI{3}{\percent}/\SI{2}{mm} was at least \SI{96.39}{\percent} and \SI{90.08}{\percent}, and every plan passed both tests. With 55 to 249 min per clinical plan on a CPU workstation, the check can be performed for each patient before the first treatment fraction.

\bibliographystyle{iopharvard}
\bibliography{references}

\end{document}